\documentclass[twocolumn,prb,aps,superscriptaddress]{revtex4}

\usepackage{graphicx}
\usepackage{dcolumn}
\usepackage{amsmath}

\newcommand{\figurewidth}{8.3cm}
\newcommand{\beq}{\begin{equation}}
\newcommand{\eeq}{\end{equation}}

\begin{document}

\title[Short Title]{
Hydrogen-Helium Mixtures in the Interiors of Giant Planets}

\author{J. Vorberger}
\affiliation{Geophysical Laboratory, Carnegie Institution of Washington,
          5251 Broad Branch Road, NW, Washington, DC 20015} 
\author{I. Tamblyn}
\affiliation{Department of Physics and Atmospheric Science, Dalhousie University,
Halifax, Nova Scotia B3H 3J5, Canada} 
\author{B. Militzer}
\affiliation{Geophysical Laboratory, Carnegie Institution of Washington,
          5251 Broad Branch Road, NW, Washington, DC 20015} 
\author{S.A. Bonev}
\affiliation{Department of Physics and Atmospheric Science, Dalhousie University,
Halifax, Nova Scotia B3H 3J5, Canada} 

\begin{abstract}
Equilibrium properties of hydrogen-helium mixtures under conditions similar to
the interior
of giant gas planets are studied by means of first principle density
functional molecular dynamics simulations. We investigate the molecular 
and atomic fluid phase of hydrogen with and without the presence of helium 
for densities between $\rho=0.19$ g$\,$cm$^{-3}$ and
$\rho=0.66$ g$\,$cm$^{-3}$ and temperatures from $T=500\,$K to 
$T=8000\,\mbox{K}$. Helium 
has a crucial influence on the ionic and electronic structure of the liquid.
Hydrogen molecule bonds are 
shortened as well as strengthened  which leads to more stable hydrogen molecules 
compared to pure hydrogen for the same thermodynamic conditions. The
{\it ab initio} 
treatment of the mixture enables us to investigate the validity of the widely 
used linear mixing approximation. We find deviations of up to $8\%$ in energy and
volume from 
linear mixing at constant pressure in the region of molecular dissociation.
\end{abstract}

\date{\today}


\maketitle

\section{Introduction}
The discovery of the first extrasolar planet in 1995~\cite{mayor95} 
marked the beginning of a new era in
planetary science, which is characterized by great improvements in
observational techniques and a rapidly expanding set of known
extrasolar planets. Most of the about 200 known planets are giant gas
planets in small orbits since the primary tool for detection, radio
velocity measurement, is most sensitive for finding heavy planets
that rapidly orbit their parent star \cite{Bu05,exopl}.
From radius measurements of transient extrasolar planets, we know that most of
the discovered planets consist primarily of hydrogen and helium.
Therefore, there is a great need for accurate 
equation of state (EOS) data for these elements under giant gas planet conditions
\cite{Gu02}.
The knowledge of equilibrium properties of mixtures of hydrogen and 
helium will help to clarify questions concerning the inner structure, origin, and
evolution of such astrophysical objects. Open questions
are whether or not hydrogen and helium phase-separate inside giant planets, if a 
plasma phase transition 
\footnote{phase transition connected with dissociation and
ionization of the neutral fluid} 
under the influence of helium can be found, and if a
solid rocky core exists in Jupiter \cite{Gu02,SG04}.

The EOS of hydrogen has attracted considerable attention and a large
number of models have
been introduced to characterize hydrogen at high pressure and temperature. 
Of great use in
astrophysical calculations and planet modeling are (free energy) models operating in
the chemical picture \cite{Stevenson75,SC92,SC95,WC05,Ju01,ju03,bez04a,Be99}. In
these models, the hydrogen fluid is assumed to be composed of well-defined
chemical species like atoms, molecules, and free charged particles.
Such methods operate in
the thermodynamic limit and are capable of describing large parameter regions of
temperature and density. Further advantages of the free energy models are the 
small computational effort required to calculate the EOS and the 
explicit knowledge of all the considered contributions to 
the EOS. 
Ionization and dissociation degrees are computed by means of mass action
laws and are not subject to fluctuations due to technical issues as in
simulations.
Atoms and molecules are treated as separate elementary species 
instead of being 
considered as bound states of electrons and nuclei. This implies certain
approximations that limit the quality of these approaches. 
At sufficiently high density, the definition of atoms and molecules becomes
imprecise as the lifetime of such objects decreases rapidly with density 
and mean distances between nuclei and electrons 
become comparable to bond lengths.

In this paper, special emphasis will be placed on testing the accuracy of the 
linear mixing approximation, which is often applied
in free energy models to calculate the EOS of mixtures of different chemical
species such as hydrogen and helium. A similar approach can be used to 
characterize mixtures of 
hydrogen atoms and molecules \cite{SC95,Ro98}.
This approximation allows the calculation of
thermodynamic variables of mixtures by a simple linear superposition of
properties of pure substances. Linear mixing is a useful assumption to make if
reliable experimental or theoretical data are only available for pure substances, 
or for cases where it has been shown that particles interact only weakly.

To avoid the shortcomings of chemical models, first principle
calculations can be applied. Such methods work in the physical picture and 
treat electrons and nuclei as elementary particles interacting via the Coulomb
potential. Quantum theory then describes the effects leading to the
formation of atoms or molecules and their statistics. 
For hydrogen, there have been great efforts to study the equilibrium
properties by means of density functional theory (DFT)
\cite{KS65,Klepeis1991, mazin95},
DFT-molecular dynamics (DFT-MD) \cite{CP85,winis04,ko95}, 
DFT - hypernetted chain equation combination (DFT-HNC) \cite{DW02,dw82,xu98}, 
path integral Monte Carlo (PIMC) \cite{Ma96,PC94}, 
couple electron-ion Monte Carlo (QMC) \cite{Pierleoni04}, and
Green's function theory \cite{tsch,vorb04}. 

Questions addressed 
include the problem of the hydrogen Hugoniot \cite{MC00,Mi01,bez04,bonev2004} and
helium Hugoniot \cite{Mi06}, the nature of the transition in hydrogen from a 
molecular to an atomic state 
\cite{Scandolo2003,pfaffenzeller97,Ma96}, the melting line 
of hydrogen \cite{bonevNature}, the different molecular
solid phases \cite{natoli95,johnson00,mazin95,ki00}, and the atomic 
solid (metallic Wigner crystal) proposed to be
found at very high pressures \cite{Wi35,Jo96,MG05}. 
Although DFT-MD is primarily an electronic groundstate method, it can be readily
applied to describe dense solid and fluid hydrogen and helium
at conditions relevant to giant gas planets because the electrons in such 
systems are either chemically bound or highly degenerate.

To our knowledge, there are only a few first
principle calculations dealing with the question of the helium influence on the
hydrogen EOS. Klepeis {\it et al.}\cite{Klepeis1991}
made predictions concerning the hydrogen-helium phase
separation but the DFT method applied in that paper is not suitable to treat 
the high temperature liquid found inside giant gas planets since only lattices
could be discussed.
The first DFT-MD calculations, performed by Pfaffenzeller {\it et al.}
\cite{Pfaffenzeller1995}, lead to more reasonable values for hydrogen-helium
demixing. However, their simulations were performed using  Car-Parinello MD 
(CP-MD). 
The high temperature region ($T\ge 15000$K) where partial ionization occurs was considered 
by Militzer \cite{hheburkhard}. 

Here we present new results concerning the hydrogen and hydrogen-helium
EOS for conditions inside giant gas planets as derived from first principle
DFT-MD. We use Born Oppenheimer MD (BO-MD) in order to ensure well 
converged electronic wavefunctions at every step. 
The density and temperature values chosen cover the region of molecular
dissociation where we expect corrections to the linear mixing approximation 
to be most significant.

We continue with Section \ref{method} which contains details of our
computational method. Results for the hydrogen EOS are presented in
Section  \ref{purehyd} and compared to EOS data from a variety of other
approaches. Furthermore, the ionic and electronic structures of the 
hydrogen fluid and their dependence on temperature and density are 
investigated as well. The EOS and properties of hydrogen-helium
mixtures are studied in Subsection \ref{hydhel}.  The focus there is on 
understanding how
the presence of helium influences the stability of hydrogen molecules and
the electronic structure, as well as on determining excess mixing quantities. 
Finally, the validity of the linear mixing approximation is examined in 
Subsection \ref{mixsec} and Section \ref{summ} provides a summary of our results and conclusions.

\section{Method}\label{method}
We use first principle DFT-MD within the physical picture to describe 
hydrogen-helium mixtures under giant gas planet conditions. This
means that protons as well as helium nuclei are treated classically.
Nuclei and electrons interact via a Coulomb potential. 
Since $T \ll T_F$, where $T_F$ is the Fermi-temperature,
for all densities and temperatures found inside a typical giant
gas planet, we employ ground state density functional theory to describe
the electrons in the Coulomb field of the ions. The ions have sufficiently 
large mass to be treated as
classical particles and their properties
described well by means of molecular dynamics simulations. We employ
the Born-Oppenheimer approximation to decouple the dynamics of electrons and 
ions. The electrons thus respond instantaneously to the ionic 
motion and the electronic wave functions are converged at every ionic time 
step. Compared to 
Car-Parinello MD, this reliably keeps the electrons in their ground state 
without the necessity to use an artificial electron thermostat for systems 
with a small band gap. 

The calculations presented in this article were carried out
with the CPMD package \cite{CPMD}. All MD results were obtained within the 
NVT ensemble. A
Nos\`e-Hoover thermostat was applied to adjust the system temperature.
The thermostat was tuned to the first vibration mode of the hydrogen molecule
($4400\,\mbox{cm}^{-1}$). All DFT-MD simulations were with 128 
electrons in super cells with periodic
boundary conditions and convergence tests were performed with larger cells. 
An ionic time step of 
$\Delta t=16\,\mbox{a.u.}$ ($1\,\mbox{a.u.}=0.0242\,\mbox{fs}$) was used 
throughout, 
however we found that $\Delta t=32\,\mbox{a.u.}$ is already
sufficient for $r_s \ge 1.86$ in the hydrogen-helium mixtures.
($r_s=3/\{(4\pi n)^{1/3}a_B\}$, $r_s$ is the Wigner-Seitz radius, $n$ the 
number density of electrons per unit volume). 
All simulations were run for at least 2 ps and  
for the calculation of thermodynamic averages, such as pressure and energy, 
an initial timespan of at least $0.1\,\mbox{ps}$ was not considered to 
allow the system to equilibrate.

The DFT calculations were performed with plane 
waves up to a cutoff energy of $35-50\,\mbox{Ha}$, the
Perdew-Burke-Ernzerhof GGA approximation \cite{PBE} for the 
exchange-correlation energy, and $\Gamma$-point 
sampling of the Brillouin zone. We used local Troullier-Martins 
norm conserving pseudopotentials \cite{TM91,fhi}. 
The
pseudopotentials were tested for transferability and for reproducing 
the bond length 
and groundstate energy of single hydrogen molecules as well as of helium 
dimers. 

For each density, the simulations were started at low temperature where
the system is in a
molecular phase and the temperature was increased in steps of 
$\Delta T=500\,\mbox{K}$ in order to avoid the premature destruction of 
molecules by temperature oscillations introduced by the thermostat.

The electronic density of states (DOS) were calculated for snapshots from MD
simulations with the Abinit package
\cite{Abinit} and using a Fermi-Dirac smearing. The presented results for 
DOS and bandgaps are based on
multiple snapshots for each parameter set.

Finite size effects were tested for by carrying out simulations with supercells
containing up to $324$ electrons (plus the required neutralizing number of 
protons and helium nuclei). For densities corresponding to  
$1.86 \ge r_s \ge 1.6$ and at $T=500\,\mbox{K}$ we found the changes in 
pressure and energy to be smaller than $2\%$. 
The convergence of the Brillouin zone sampling was checked by optimizing the 
electronic density of several
MD snapshots with
$\Gamma$-point, a $2\!\times\! 2\!\times\! 2$, and a 
$4\!\times\! 4\!\times\! 4$ Monkhorst-Pack grid of 
{\bf k}-points \cite{MP76} for a
$N_e=128$ system. The Brillouin zone appears to be sufficiently small 
so that deviations between results with $1$, $8$, and $64$ ${\bf k}$-points are
below $1\%$ ($r_s=1.6$, $T=500\,\mbox{K}$). 

We also examined the significance of electronic excitations for the 
thermodynamic properties of the studied fluids. Snapshots from MD trajectories
were taken and electronic states were populated according to a Fermi 
distribution corresponding to the MD temperature. For $r_s=2.4$ and
$T=7000\,\mbox{K}$ the pressure was found to increase by $8\%$. Since
the degeneracy parameter of the electrons
increases while moving along an isentrope to the center of a giant gas 
planet, this can be considered an upper limit for finite temperature 
electronic excitation effects; the error for higher densities and lower
temperatures will be much smaller.
%
%
%
\section{Results}
Here we present {\it ab initio} results for equilibrium properties of
hydrogen and hydrogen-helium mixtures in a density region between
$0.19\,$g$\cdot$cm$^{-3}$  and $0.66\,$g$\cdot$cm$^{-3}$ ($r_s=2.4$ to
$r_s=1.6$) and for temperatures from $500\,\mbox{K}$ to
$8000\,\mbox{K}$.  This parameter region includes part of the
transition region from the  molecular to the atomic fluid state of
hydrogen and hydrogen-helium mixtures.  It is, therefore, interesting
to study not only to get insight into interior properties of giant gas
planets but also to examine molecular dissociation, the
molecular-atomic and the  insulator-metal transitions in
hydrogen. Additionally, one can consider the  influence of helium on
these transitions and properties of mixing.

\subsection{Pure Hydrogen}\label{purehyd}
\begin{figure}[!]
\includegraphics[angle=0,width=\figurewidth]{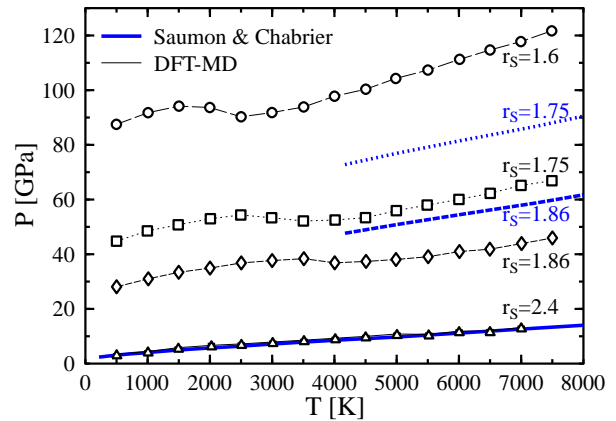}
\caption{Pressure-temperature relation for pure hydrogen  along various 
isochores in DFT-MD (this work) and according to
Saumon and Chabrier (SC)~\cite{SC92,SC95}. The isochore of SC
for $r_s=1.6$ lies out of the range of the $P$ axis. 
Errors for the DFT-MD simulations are of the order of the symbols.
 }
\label{isocha}
\end{figure}

Figure \ref{isocha} provides a summary of our hydrogen EOS calculations using
DFT-MD. 
Four different pressure isochores are shown.
At low density ($r_s=2.4$), the pressure increases monotonically with 
temperature as the character of the fluid changes smoothly from molecular to
atomic. At this density, the transition is slow 
enough with temperature so that the drop in the pressure when molecules
break (the interactions become less repulsive) is compensated by the increase 
of the kinetic contribution to the pressure.

At higher density ($r_s<2$), the 
dissociation of molecules takes place more rapidly with increasing temperature
and leads to a region of $\partial P/\partial T|_V<0$. 
At sufficiently high density, this effect dominates
over the pressure increase that results from the presence of two atoms instead of
one molecule. Furthermore, the condition $\partial P/\partial T|_V<0$ implies a negative 
thermal expansivity $\partial V/\partial T|_P<0$, while the fluid maintains 
hydrostatic stability given by $\partial P/\partial V|_T>0$.

By exhibiting a region with $\partial P/\partial T|_V<0$, fluid hydrogen shares
some properties with typical solids, where a new crystal structure with more
efficient packing appears and the pressure is lowered at fixed volume. In solid
hydrogen different transition pressures to an atomic solid have been predicted, 
above $300\,\mbox{GPa}$ \cite{Lo02,CA87} and $r_s =\sim 1.3$. This is consistent 
with the observed shift of the region of 
$\partial P/\partial T|_V<0$ to lower temperatures, as the density is increased.
It indicates a density effect on dissociation as less and less thermal energy is
needed to break up the molecular bonds. At even higher densities than shown here,
the bond length is equal or less than the mean particle distance. In this
regime the interaction of molecules and atoms becomes too strong and 
pressure dissociation/ionization occurs.

\begin{figure}[!]
\includegraphics[angle=0,width=\figurewidth]{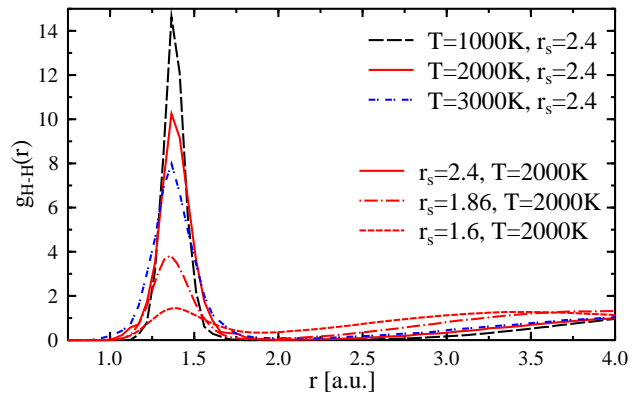}
\caption{Pair correlation functions $g(r)$ for pure hydrogen at 
different temperatures (color coded) and different densities: 
solid, dash-dotted, and dashed lines, respectively: 
$r_s=1.6 \,(0.66\,\mbox{g}\cdot\mbox{cm}^{-3}$), 
 $r_s=1.86 \,(0.42\,\mbox{g}\cdot\mbox{cm}^{-3})$, 
 $r_s=2.4 \,(0.20\,\mbox{g}\cdot\mbox{cm}^{-3})$.
 }
\label{bindist}
\end{figure}
Figure \ref{bindist} shows the dependence of pair correlation
functions (pure hydrogen at $r_s=2.4$) on the temperature. 
The first peak at $r \approx 1.40\,$a.u. indicates
the existence of hydrogen molecules 
With increasing temperature, the height of the first peak is reduced 
as molecules dissociate.
An analogue behavior can be observed by plotting the changes in $g(r)$ with
density. 
A strong decrease of the first peak with increasing density and thus a
significant lower fraction of molecules at higher densities is revealed.
In addition to the less pronounced first maximum, an overall weakening of
the short range order can be observed with increasing temperature. 

\begin{figure}[!]
\includegraphics[angle=0,width=\figurewidth]{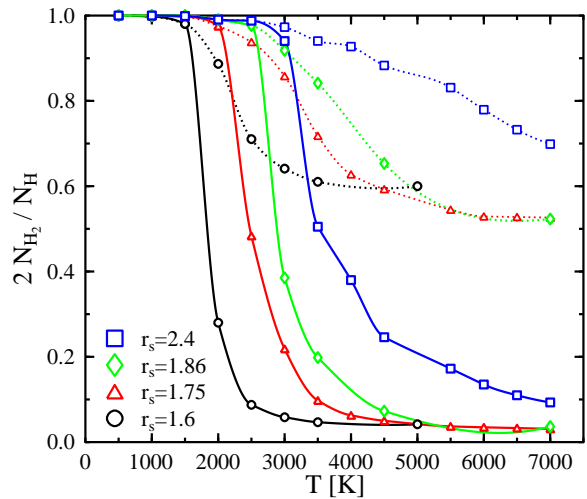}
\caption{Dissociation degree of the hydrogen molecules in pure hydrogen. The 
dissociation degree obtained by simply counting all pairs of hydrogen atoms with
distance shorter than $r_{cut}=1.8\,\mbox{a.u.}$ is plotted with dotted lines. 
Full lines take into
account the lifetime of these pairs as well ($10$ H$_2$ vibrations
at least to be counted as molecule).
 }
\label{dissoc}
\end{figure}

A more quantifiable picture of the described effects can be obtained by 
plotting the degree of dissociation, 
\begin{equation}
\alpha=\frac{2N_{H_2}}{N_H}\;,
\end{equation}
as a function of temperature (Fig. \ref{dissoc}).
Here $N_{H_2}$ is the average number of hydrogen molecules at the given density 
and temperature conditions.
$N_H$ the total number of hydrogen nuclei irrespectively of the dissociation
state. 

At higher density, fewer molecules are 
present at the same temperature as a result of pressure dissociation. 
While at low density, the dissociation proceeds gradually
with temperature, 
the curves for $r_s\le 1.75$ show a rapid drop around 
$2500\,\mbox{K}$, which is related to the $\partial P/\partial T|_V<0$ region.


The dissociation degree and the binary distribution function are nevertheless not
sufficient to draw a complete picture of the structure and dynamics in fluid 
hydrogen. The
lifetime of the molecules must also be taken into account. Figure \ref{dissoc}
shows that at $r_s=1.75$, for example, even though on average more than $50\%$
of the protons are found in paired states, the lifetime of 
these pairs is short (less than 2 H$_2$-vibrations on average); there is a 
continuous formation and destruction of
pairs of hydrogen atoms. It is therefore imprecise to classify the fluid as
either molecular-atomic or pure atomic as there is no unique criterion
for a molecule. However, the results for the EOS obtained by our simulations
do not depend on the number of molecules or atoms but only on temperature and 
density.

\begin{figure}[!]
\includegraphics[angle=0,width=\figurewidth]{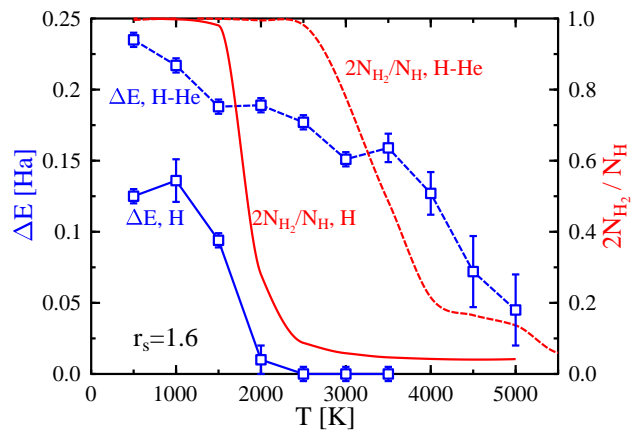}
\caption{Bandgap calculated in GGA for fluid hydrogen and hydrogen-helium
mixture (blue curves) as well as dissociation degree for the two systems
(red curves). The density corresponds to $r_s=1.6$. 
The dissociation degree is determined by taking into account lifetime effects.}
\label{gap}
\end{figure}

In addition to the dissociation of hydrogen 
molecules, of interest are the changes in the electronic structure
taking place in the same parameter region. As the system 
becomes denser or the temperature is raised (still $T\ll T_F$), 
the interactions
between the molecules in the fluid become stronger, and the formerly well-bound
electrons become delocalized. This is associated with a strong increase in the
electrical conductivity and is usually referred to as metallization \cite{We96}.
The effect can be seen in the electronic density of states (DOS), namely the 
band gap, as shown in
Fig. \ref{gap}. We calculated the Kohn-Sham eigenvalues in GGA for several
snapshots and estimated the bandgap in fluid hydrogen along the MD trajectory.
For $r_s=1.6$, a closing of the bandgap can be observed around $T=2000K$. 
This  means that a metallic-like state may have been formed. 
Furthermore, as indicated by the red
lines in Fig. \ref{gap}, the degree of 
dissociation incorporating lifetime effects decreases strongly around the same 
temperature. To distinguish this
degree of dissociation we do not consider proton pairs with less than 10
H$_{2}$ vibrations ($t=10\times 7.6\,\mbox{fs}$) as being molecules. The closing of 
the bandgap and dissociation of hydrogen molecules happen at the same time.

It is well known that the GGA underestimates the bandgap. 
More sophisticated calculations \cite{gi84} (for a $H_2$ solid) give a bandgap 
of $\sim 0.6\,$ Ha at $T=300\,\mbox{K}$. The value we obtain is $4$ times smaller.
Hood and Galli \cite{HoodGalli} compared values obtained with DFT (GGA) to 
quantum Monte Carlo (QMC) gaps for liquid deuterium. For $T=3000\,\mbox{K}$ and
$r_s=1.6$, they found the QMC gap twice as large as the DFT gap.
The actual temperature of metallization is thus 
somewhat higher. 
To improve the description of the electronic properties of the fluid, one
needs to use a more accurate method than DFT-GGA, which is beyond the 
scope of this article.
Still, it is worth noting that we find a continuous transition from an 
insulating to a conducting 
state, as determined by the closing of the gap.
We do not observe  molecules in the conducting phase as 
found by Weir {\it et al.} \cite{We96} or Johnson {\it et al.} \cite{johnson00}.

While there is a general agreement about dissociation, ionic, and electronic 
structural changes throughout various
papers, this agreement is only qualitative. As can be seen in Fig.
\ref{isocha}, different methods give very different results for the EOS of dense
fluid hydrogen. Whereas for the lowest density shown in Fig. \ref{isocha}, 
the
agreement between the free energy model of Saumon and Chabrier \cite{SC92,SC95} 
and our results is reasonable, deviations up to $20\%$ (at $r_s=1.86$) and even 
$24\%$
(at $r_s=1.75$ and above) can be found for higher densities. The free energy 
model overestimates the pressure considerably. The degrees of dissociation 
calculated with the free energy
method show a significantly higher fraction of molecules than we find in our 
simulations. However, even a linear extrapolation of our pressure results of the 
molecular phase to higher temperatures 
(a linear scaling very similar to the one at $r_s=2.4$ is assumed) only reduces 
the discrepancy with the result of
Saumon and Chabrier but can not eliminate the difference completely. Deviations 
of this order may signify completely different physics inside giant gas planets 
and it is of great importance to discuss the discrepancies.

\begin{figure}[!]
\includegraphics[angle=0,width=\figurewidth]{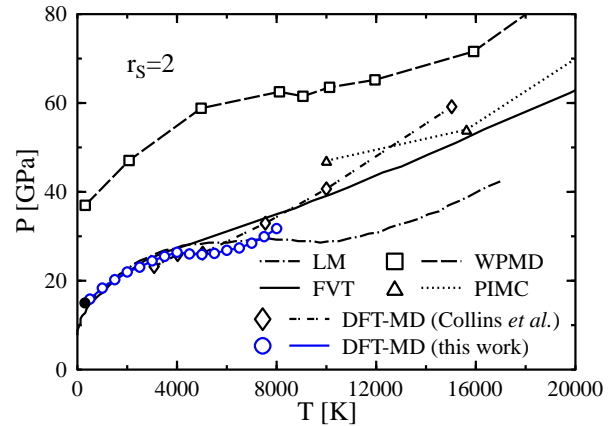}
\caption{Pressure-temperature relation for a single isochore of hydrogen 
with different methods for $r_s=2$. Our DFT-MD results are shown in blue 
with circles.
FVT by Juranek
{\it et al.} \cite{Ju00} (solid black line), PIMC by Militzer {\it et al.} \cite{MC00}
(dotted line, triangles), WPMD by Knaup {\it et al.} \cite{KRT98} (dashed line,
squares), DFT-MD by Collins {\it et al.} \cite{Co95} (dashed, diamonds), LM by Ross
\cite{Ro98} (long-short dashed), and this works DFT-MD results (blue, dots). The
black dot indicates the pressure of a $H_2$-solid at $T=300\,\mbox{K}$ 
\cite{Lou96}. The
error bars of simulation results are within the size of the markers.
 }
\label{isoch}
\end{figure}
Comparisons with different first principle calculations can help resolve this 
issue, as agreement between different independent {\it ab initio} methods would be a 
strong 
indication for correct results. Figure \ref{isoch} provides such a comparison. In
addition to our results, PIMC data \cite{MC00}, wave packet
molecular dynamic (WPMD) results \cite{KRT98,Kn02,Kna02} and older DFT-MD points
\cite{Co95} are shown. Furthermore, the isochores of two different models in the
chemical picture are added: fluid variational theory (FVT) \cite{Ju00} as well as
the linear mixing (LM) model \cite{Ro98}. They start from a mixture of atoms and
molecules and their (Lennard-Jones type) interactions to minimize the free energy
with respect to the fraction of the constituents. The isochore provided by
WPMD deviates from the other ones by more than a factor of two at lower
temperatures and by $25\%$ at the highest temperatures shown here. 
PIMC is not a ground state method and is more capable of determining the 
EOS at 
higher temperatures. Information about fluid hydrogen or even about dissociation
of hydrogen molecules can not be obtained. For higher temperatures, PIMC results
lie between DFT-MD and FVT data.
Better agreement is achieved between the two DFT-MD
methods, FVT and LM. In the region with temperatures less than 
$10000\,\mbox{K}$ some features of the curves are still different. 
The first principle simulations 
show a region with a reduced or even slightly negative slope 
(smooth transition from a purely molecular to an atomic fluid)
around $5000\,\mbox{K}$. Such a behavior is absent in the FVT model. The LM
method, on the other hand, shows a similar feature but on a much wider
temperature range. 
Thus, there is
no unique picture of the EOS of dense fluid hydrogen. Deviations 
between the results of different first principle methods
are related to the treatment of the electrons. Inconsistencies between 
the chemical 
picture, as a basis for FVT and LM, and the physical picture, as a foundation 
of 
DFT-MD or PIMC, contribute to the non-unique description. However that
may be, DFT-MD provides reproducible results (the two DFT-MD studies were
performed with 
different codes) and FVT seems to be reliable in the molecular fluid phase.

\begin{figure}[!]
\includegraphics[angle=0,width=\figurewidth]{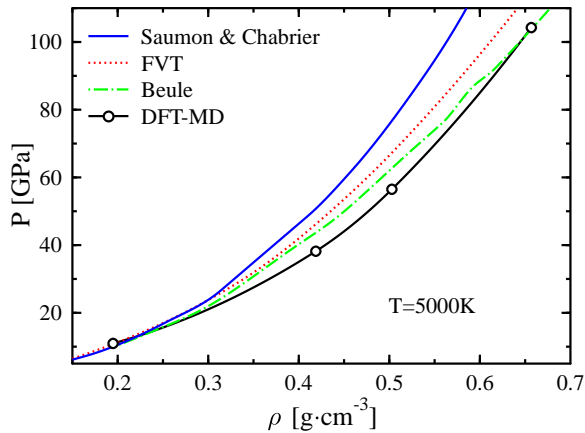}
\caption{The $5000\,\mbox{K}$ pressure isotherm of hydrogen computed 
with various methods: 
this works DFT-MD (black with circles), FVT (red) \cite{Ju01}, 
Beule (FVT+ionization, green) \cite{Be99}, 
Saumon and Chabrier (blue solid) \cite{SC95}. 
HNC-MAL \cite{bez04a} gives similar results as FVT.
 }
\label{isoth}
\end{figure}
Lastly, we consider more closely the intermediate density 
range at a
temperature typical for the interior of Jupiter. Figure \ref{isoth} shows
such an isotherm for $T=5000\,\mbox{K}$. Our result gives the lowest pressure.
Saumon and Chabrier's EOS predicts a pressure up to $20\%$ higher as stated 
above. The blue curve, as well as the green one, show a plasma phase 
transition 
(PPT). According to Saumon and Chabrier the PPT is expected at a density
of approximately $\rho=1\,$g$\cdot$cm$^{-3}$ ($P=200\,\mbox{GPa}$).
The FVT results with \cite{Be99} or without \cite{Ju01} extension to ionized 
plasmas gives an
isotherm right between the two above mentioned results. These methods
predict a PPT at slightly smaller density of $\rho=0.8\,$g$\cdot$cm$^{-3}$ and
$P=100\,\mbox{GPa}$).
At both densities, neither we nor Weir {\it et al.}
\cite{We96} have evidence for such a PPT. Instead, a continuous transition from
a molecular to an atomic state takes place at lower temperatures. 
It is nevertheless remarkable that the inclusion of ionization 
reduces the pressure and gives better agreement. From our simulation we have 
no information about possible intermediate ionized states of hydrogen. 

\subsection{Hydrogen-Helium Mixtures}\label{hydhel}
So far we have studied pure hydrogen under giant gas planet conditions. A 
further degree of freedom is added if one considers a mixture of hydrogen and 
helium. Helium,
even in small fractions, changes the EOS significantly. Helium has an influence
on the formation and dissociation of hydrogen molecules, and it changes the ionic
structure of the liquid as well as the electronic properties. The transition
from a molecular state into an atomic state may be displaced or its character
changed. Further, hydrogen and helium have been predicted to phase-separate in giant
planet interiors.
\cite{Stevenson77a,Stevenson77b,Stevenson75}.

Models in the chemical picture use the linear mixing
rule to add hydrogen and helium portions to the EOS \cite{SCapj92}. 
Contributions from the entropy of mixing are ignored and all
the interactions between the two subsystems are left out. 
First principle calculations include all these effects since a mixture of the two
fluids can be simulated directly.
The demixing line was calculated by classical Monte Carlo \cite{Hu85}, by 
ground state DFT calculations \cite{Klepeis1991}, and by Car Parinello MD 
\cite{Pfaffenzeller1995}.

Here, we primarily present results for a hydrogen-helium mixture at a mixing
ratio of $x=0.5$. The mixing ratio is defined as
\begin{equation} 
x=\frac{2N_{He}}{(2N_{He}+N_{H})}\;,
\label{xfrac}
\end{equation}
where $N_{H}$  and $N_{He}$ are the number of hydrogen and helium nuclei per unit
volume. This definition weights the species according to the number of electrons
that they contribute to the system. The corresponding Wigner-Seitz radius is
computed from the total number of electrons. For many simulations, a mixing ratio
of $x=0.5$ was chosen so that large interaction effects between the two species
could be observed.

\begin{figure}[!]
\includegraphics[angle=0,width=\figurewidth]{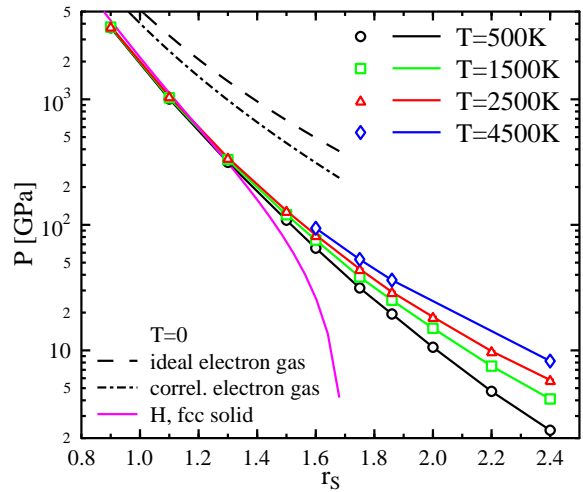}
\caption{Pressure isotherms of a mixture of hydrogen and helium 
(x=0.5) for different temperatures. High density limiting results for hydrogen
and an electron gas are shown additionally.
 }
\label{isothhe}
\end{figure}
Figure \ref{isothhe} shows the pressure for a number of isotherms for a 
hydrogen-helium mixture. 
The maximum density shown here corresponds approximately to conditions in the 
center of Jupiter ($r_s=0.9\,,\; \rho=3.6\,\mbox{g}\,\mbox{cm}^{-3}$). It is 
demonstrated that temperature is not important for higher densities. 
since all the
isotherms merge into the one with the lowest temperature. At the highest 
densities shown here the temperature contribution of the ions to the 
pressure is approximately $5\%$. 
The ions are strongly coupled and their interaction contribution to the EOS is
of the order of $30\%$. The rest of the deviation from the ideal degenerate
system is given by nonidealities in the electron gas and interactions between
electrons and ions.
For even higher densities the electronic contributions will become even more
important since they rise
with density as $n^{5/3}$ and thus faster than any other contribution. 

\begin{figure}[!]
\includegraphics[angle=0,width=\figurewidth]{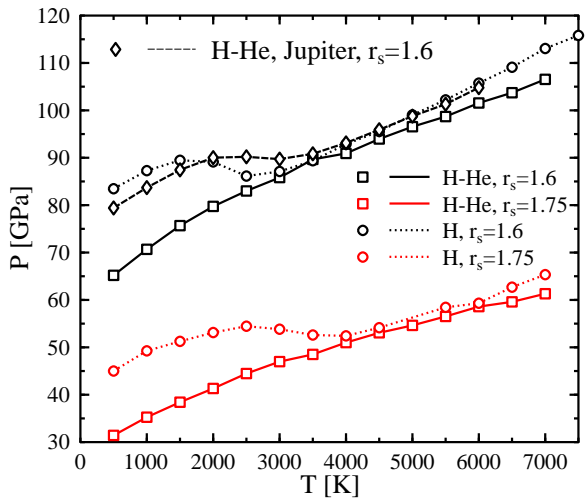}
\caption{Pressure isochores for a mixture of hydrogen and helium (x=0.5, 
solid lines) and for pure hydrogen (dotted lines) at two different densities. 
For $r_s=1.6$ an additional curve for Jupiter's helium ratio (x=0.14) 
was added (back dashed line).
 }
\label{isochhe}
\end{figure}
The role of helium for the EOS of the mixture can be studied in
Fig. \ref{isochhe}. The pressure is slightly lowered over the
whole temperature range, but more important is the fact that the 
region with negative $\partial P/\partial T|_V$ has vanished (for $x=0.5$). 
The additional curve for Jupiter's hydrogen-helium mixing ratio (x=0.14)
shows an intermediate step where the region with negative slope of the pressure
does still exist but is reduced in markedness and in the temperature range of
existence. 
More helium in the mixture means a shift of the negative slope to higher
temperatures and a decrease of the depth of the minimum. 
For  very high temperatures, the EOS of pure hydrogen and mixtures show a 
very similar behavior. Helium has a more significant influence on the pressure 
at lower temperatures. 

\begin{figure}[!]
\includegraphics[angle=0,width=\figurewidth]{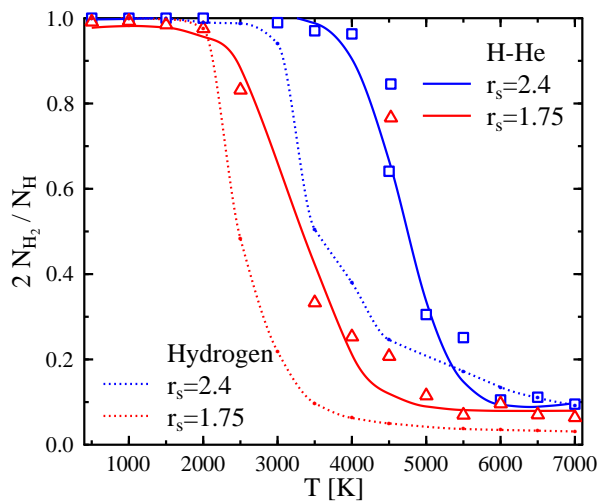}
\caption{Comparison of dissociation degrees (incorporating life time effects)
in pure hydrogen and in hydrogen-helium (x=0.5). The solid lines are nonlinear
fits to the data points.
 }
\label{dissoc_hyhe}
\end{figure}
\begin{figure}[!]
\includegraphics[angle=0,width=\figurewidth]{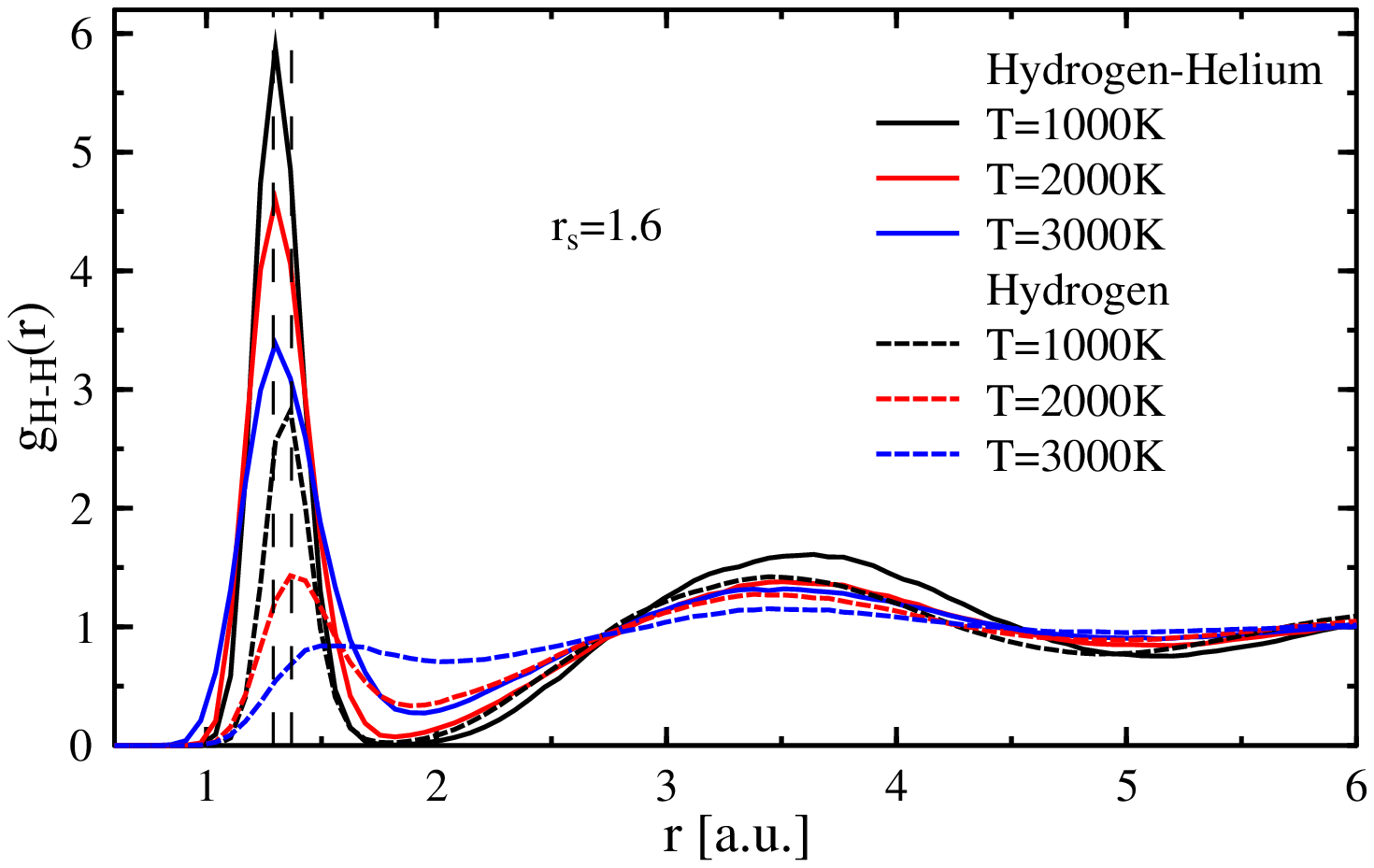}
\caption{Pair correlation functions for hydrogen (dashed lines) and 
hydrogen-helium (solid lines) at different temperatures.
The order of the curves from top to bottom is the same as in the legend. The
vertical thin dashed lines indicate the location of the first peak for hydrogen
and hydrogen-helium, respectively.}
\label{bindisthyhe}
\includegraphics[angle=0,width=\figurewidth]{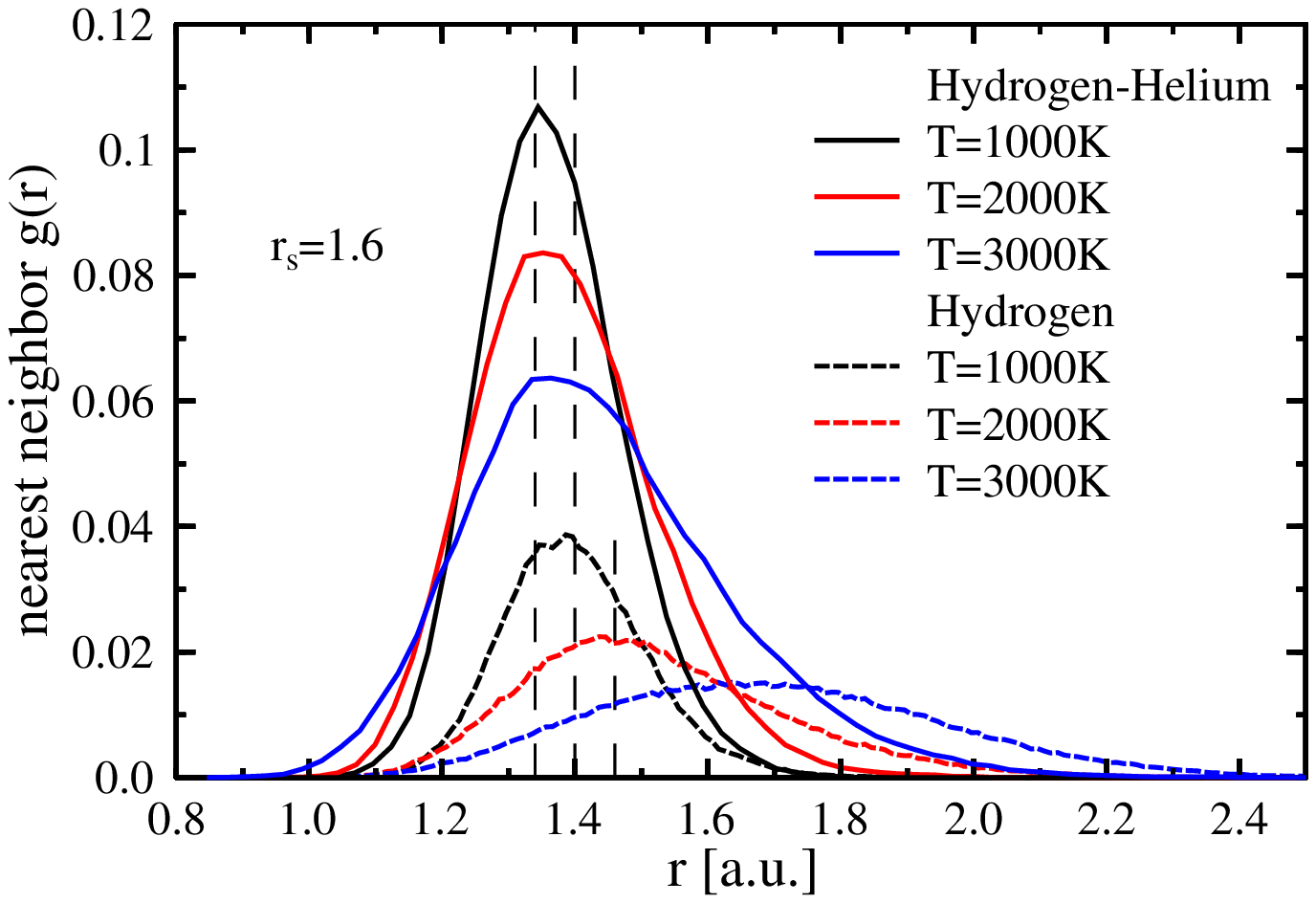}
\caption{1st neighbor distribution for hydrogen (dashed lines) and 
hydrogen-helium (solid lines) at different temperatures. 
The order of the curves from top to bottom is the same as in the legend. The
vertical thin dashed lines indicate the location of the peak for hydrogen
and hydrogen-helium, respectively.}
\label{neighborhyhe}
\end{figure}

These two different regions can clearly be assigned to different dissociation
regimes (see Fig. \ref{dissoc_hyhe}). The transition from a 
molecular phase to an atomic phase, while
still smooth, takes place at lower temperature and over a
shorter range of temperature in hydrogen than in the mixture. This relatively
rapid change in the microstructure of the fluid is the reason for the drop in the
pressure. The vanishing of the molecules in hydrogen and their extended
existence in the mixture can be confirmed with the help of pair correlation
functions in Fig. \ref{bindisthyhe}. The molecular peak (first peak) drops 
considerably faster in pure hydrogen. This behavior and the higher peak in the
mixture give clear evidence for molecules at the highest temperatures shown
here. If we take the position of the first maximum as a measure for the mean bond
length of the hydrogen molecule, we obtain a value of $\langle d\rangle=1.37$  a$_0$ for
pure hydrogen. In the mixture (with ratio of $x=0.5$) this value changes 
to $\langle d\rangle=1.29$ a$_0$ which means a shortening of the bond by
$6\%$. The same can be obtained by means of nearest neighbor distributions as in
Fig. \ref{neighborhyhe}. The first neighbor distribution considers the nearest
neighbor only and effects of particles farther away are removed from the curve.
Bond lengths obtained from Fig. \ref{neighborhyhe} are slightly bigger.
In addition, a shift of the bond length in hydrogen from 
$1000\,\mbox{K}$ to $2000\,\mbox{K}$
is revealed. The reduction of the bond length of $6\%$ is confirmed.
The latter value is in rather good agreement with data by Pfaffenzeller 
{\it et al.} \cite{Pfaffenzeller1995}. The same conclusion is derived when 
comparing pair correlation functions for pure hydrogen and hydrogen-helium mixtures 
at constant pressure instead at constant electronic density.

Figure \ref{dissoc_hyhe} describes the role of the (electronic) density during the 
process of dissociation. Helium stabilizes the molecules. This is due to 
the higher charge of $Z=2$ of the helium nuclei. The intramolecular bonds
depend strongly on the electronic behavior and the space available. If electronic
wavefunctions between different molecules start to overlap, in other words, when
Fermi statistic becomes important for the electrons of the system as a whole, and
when the distances between the particles become so small that
interactions between the molecules are no longer weak, the electrons are
forced to delocalize to obey the Pauli exclusion principle and bonding becomes impossible. 
Helium under giant gas planet conditions, in
atomic form, binds two electrons closely. The rest of the hydrogen atoms and 
electrons are affected less by
density and temperature and the molecules remain stable over a wider range. 
The helium influence is thus two-part. First, its stronger Coulomb attraction 
binds electrons. Second, (as a consequence) it influences the many particle 
state of the electrons.

\begin{figure}[!]
\includegraphics[angle=0,width=\figurewidth]{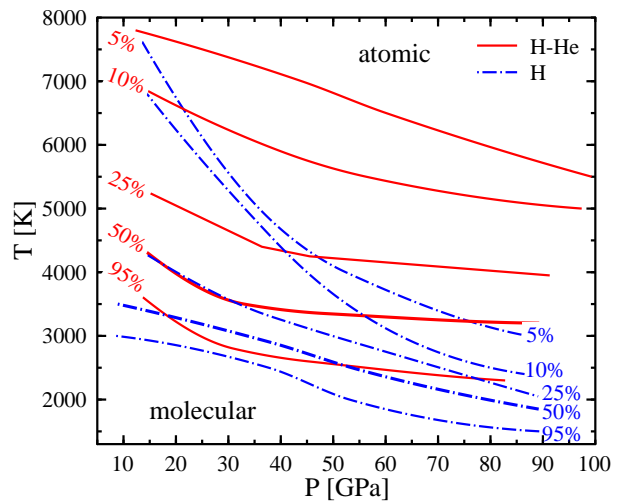}
\caption{Temperature-pressure-plane showing estimated lines of constant dissociation degree
in pure hydrogen and in a hydrogen-helium mixture (x=0.5). The percentages give 
the fraction of hydrogen atoms bound in molecules. 
 }
\label{plane}
\end{figure}
A phase diagram for molecular and atomic hydrogen and hydrogen-helium is provided
in Fig. \ref{plane}. Here, parameter regions for purely molecular, purely
atomic as well as intermediate phases are shown. 
The diagram shows
the increasing differences between hydrogen and the mixture with
increasing pressure (density) and the huge differences 
(especially at high pressure) in the
rate of the transition from a molecular to an atomic state. The transition
region (from $95\%$ to $5\%$) has nearly the same size for small pressures in
pure hydrogen and in the mixture. At the other end of the pressure scale hydrogen
changes from molecular to atomic over only $1500\,\mbox{K}$. In the mixture the
changes are more moderate. At the pressures considered here, pressure
dissociation is suppressed since the slope of the lines with constant dissociation
degree is small for higher pressures.
A good zeroth
approximation for the behavior of the mixture is given by taking into account the
hydrogen density only for the dissociation process. In this way, $r_s=1.6$ for a
$50\%$ mixture of hydrogen and helium would correspond to $r_s=2.02$ in pure 
hydrogen. This estimate works quite well for the $5\%$ line, for instance.

Band structure and electronic density of states are affected also by helium.
The inner regions of Jupiter are believed to be made of
helium rich metallic hydrogen \cite{Gu02}. The conditions under which the mixture
becomes metallic strongly depends on the amount of helium. A
comparison of the (GGA) bandgaps in pure hydrogen and a hydrogen-helium mixture
($x=0.5$) is shown in Fig. \ref{gap}. And while the bandgap in pure hydrogen 
goes to zero at relatively low temperatures, the gap in the mixture remains open
over the whole temperature region shown. This can be traced back to the charge of
the helium nucleus which shifts part of the Kohn-Sham eigenvalues to lower
energies and thus increases the gap.
\begin{figure}[!]
\includegraphics[angle=0,width=\figurewidth]{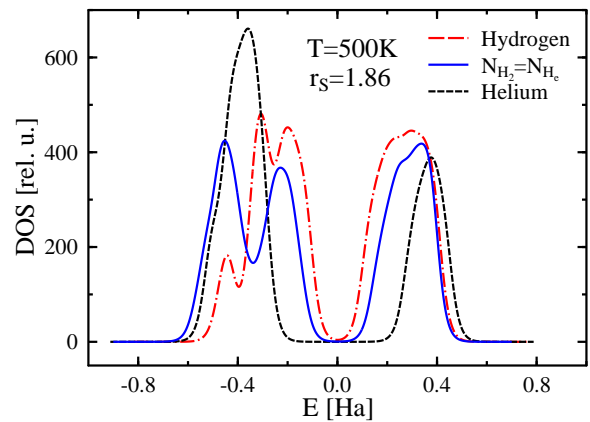}
\caption{Electronic density of states for pure fluid hydrogen, pure fluid helium,
and for a mixture of hydrogen and helium (x=0.5) at fixed $T=500\,\mbox{K}$ and
$r_s=1.86$. The pressure is approximately $30\,$GPa. Curves shifted to agree at
the Fermi energy (0 Ha) which is taken in the middle between HOMO and LUMO.
}
\label{doshhe}
\end{figure}
The change in the electronic DOS from a helium system to a mixture to a pure 
hydrogen system is shown in Fig. \ref{doshhe}. The black peak indicates atoms 
in fluid helium. The red curve shows mainly molecules in hydrogen and there are
peaks resulting from intermolecular interaction, too. The blue curve for the 
mixture is a superposition of the ones for
pure systems, namely a helium peak to the left and a hydrogen molecule peak on
the right hand side. The bands on the right hand side are empty since the curves are
normalized so that the Fermi energy is at $0$ Ha. The position of the edges of 
the bands at positive and negative energy
strongly depends on the amount of helium in the fluid and therefore the width of
the bandgap depends on the helium amount.

\begin{figure}[!]
\includegraphics[angle=0,width=\figurewidth]{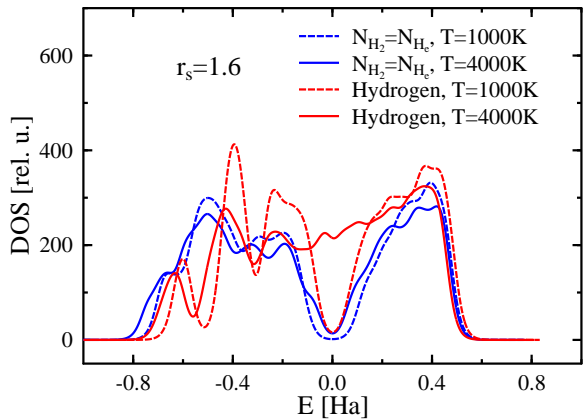}
\caption{Electronic density of states for pure fluid hydrogen and for a mixture 
of hydrogen and helium (x=0.5) as function of temperature. The pressure is 
approximately $100\,$GPa.
 }
\label{dos2hhe}
\end{figure}
The dependence of the DOS on temperature is demonstrated in Fig.
\ref{dos2hhe}. We observe a similar effect as Scandolo did \cite{Scandolo2003}
although the transition from a mainly molecular liquid to an atomic fluid
is smooth in our calculations. Due to the larger initial bandgap in the
hydrogen-helium mixture and due to the effect of the helium described
above, the bandgap for this systems remains, even at the highest temperature shown. 
Conversely, the gap in the pure system has completely closed
at $T=4000\,\mbox{K}$.
\subsection{Thermodynamic Properties of Mixtures}\label{mixsec}
We will test the validity of a common approximation used in
determining EOS's of mixtures. Most of the effort has been put into the
determination of the EOS of pure systems. By ignoring the exact nature of the
interaction between the pure phases, one can construct the EOS of any
mixture of the original phases with the help of the linear mixing (LM)
approximation
\begin{equation}
Y_{\mbox{LM}}=(1-x)Y_{\mbox{H}}+xY_{\mbox{He}}\;,
\end{equation}
with $x$ according to Eq. (\ref{xfrac}) being the fraction of helium in the mixture and 
$Y$ being a  thermodynamic variable such as volume, pressure, or internal energy.
For the free energy, an additional term describing the entropy of mixing must be
included \cite{Ro98}. Linear mixing may be performed at constant chemical potential,
at constant volume, or at constant pressure. For calculations of the
internal structure of giant gas planets, mixing under constant pressure is the
most important.
In all cases, the deviation from LM can be calculated as
\begin{equation}
\Delta Y_{\mbox{mix}}(x)=Y(x)-Y_{\mbox{LM}}(x)\;,
\end{equation}
where $Y(x)$ is the value obtained by DFT-MD for mixing fraction $x$ and
$Y_{\mbox{LM}}$
is the LM value computed from independent simulations results 
for pure hydrogen and pure helium.

In this way it is assumed that the potential between particles of
two different species can be written as an arithmetic average
over the interactions in the pure systems \cite{Ro98}. This of course works
for weak correlations only. The advantage is that one
does not need to know exactly the interaction between, e.g., hydrogen and helium. 
This would be necessary for models in the chemical picture. Therefore, this
approximation is used mainly by chemical models 
for the description of hydrogen-helium mixtures \cite{Be99,Ju00,SC92,SC95} or 
even mixtures of atoms and molecules of hydrogen \cite{Ro98}. The
error introduced is difficult to quantify. 

First principle calculations are able to verify the assumptions for LM and the
validity of the approximation since these
methods rely on the more fundamental Coulomb interaction, do not need to assume
different interaction potentials between different species, and can thus simulate
mixtures directly. There have been some investigations concerning the validity of
LM by classical MC and integral equation techniques for classical binary liquids
not including molecules
\cite{Og93,DeW96,Rosen96,DeW03}. In these cases, the deviations from LM found 
are of the order of $1\%$ and below. 

The first 
PIMC calculations for hydrogen-helium mixtures 
found deviations from LM at constant volume of up to $12\%$ for temperatures 
between $15000$ and $60000\,\mbox{K}$ and giant gas planet 
densities ($r_s=1.86$) 
\cite{hheburkhard}. This gives reason to expect that linear mixing 
might give a slightly falsified picture of the EOS of  
hydrogen-helium at lower temperatures, too. Again, since the conditions for 
phase separation depend strongly on small changes in the EOS 
(and thus from deviations from LM) it is
crucial to investigate LM \cite{DeW93b}. 
\begin{figure}[!]
\includegraphics[angle=0,width=\figurewidth]{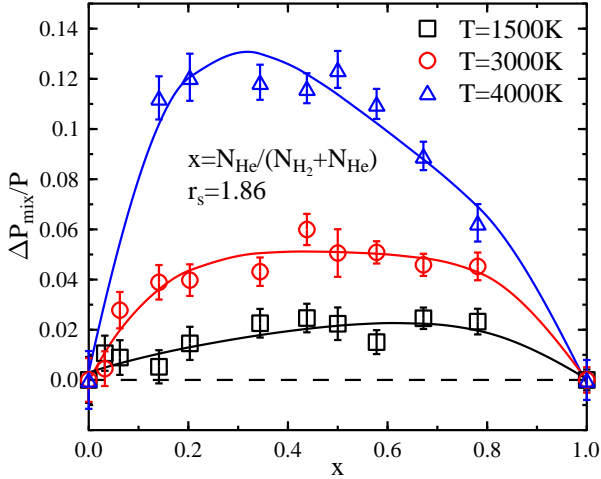}
\caption{Mixing error in the pressure due to the linear mixing approximation 
at constant volume for
various temperatures as function of the mixing ratio. The electronic density 
is $r_s=1.86$. This corresponds to
$\varrho=0.42\,$g$\cdot$cm$^{-3}$ for pure H and $\varrho=1.66\,$g$\cdot$cm$^{-3}$ 
for pure He (pressure
between $10\,$GPa and $40\,$GPa). The symbols represent calculated values, the lines
were obtained with a polynomial fit of fourth order.
 }
\label{mixv2}
\end{figure}
The advantage of first principle calculations is
the correct treatment of the degenerate electrons and bound states, which is
missing in the classical simulations.

The error introduced due to LM at constant volume (constant electronic density) 
as observed within DFT-MD is plotted in Fig.
\ref{mixv2}. Similar to the findings of other authors \cite{DeW03,hheburkhard},
the error is positive. As expected, the deviation in the pressure from the LM value is
biggest for $x=0.5$. Furthermore, increasing temperature causes an increase in 
the LM
error of up to $12\%$ for the highest temperature plotted in Fig. \ref{mixv2}. The
deviation from LM for a Jupiter like mixing ratio of $x\approx 0.14$ ranges from
around zero ($T=500\,\mbox{K}$) up to $10\%$. The temperature inside Jupiter for this
density is according to Saumon and Chabrier \cite{SC92,SC95} of the order of
$5000\,\mbox{K}$. This means that one can expect a deviation of approximately $10\%$ of
the true EOS from the one calculated with LM.

\begin{figure}[!]
\includegraphics[angle=0,width=\figurewidth]{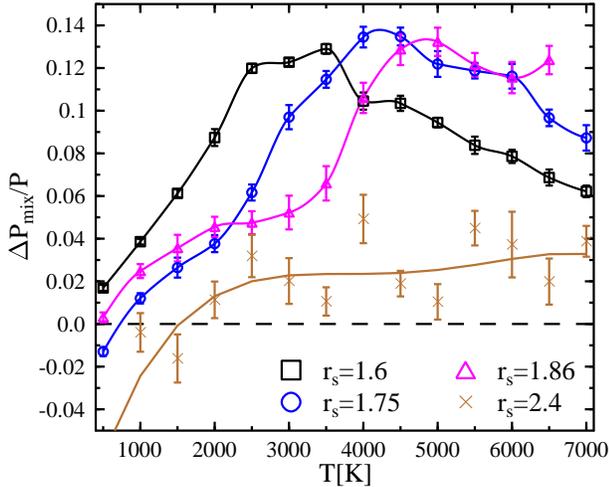}
\caption{Mixing error in the pressure due to the linear mixing approximation at 
constant volume for various densities as function of temperature. The mixing
ratio is $x=0.5$. The $r_s=2.4$ data points suffer from simulation noise and the
curve was obtained by least square fitting a third degree polynomial.
 }
\label{mixv1}
\end{figure}
The dependence of the error maximum at $x=0.5$ from density and temperature is
shown in Fig. \ref{mixv1}. The curve for the smallest density shown here 
($r_s=2.4$) gives reason to conclude that LM is a good approximation for the pure
molecular phase of hydrogen and helium as found at this density over a wide
temperature range. With increasing density the deviations from
LM start to grow as well and corrections to the pressure become 
significant for smaller temperatures. $5\%$ error is reached around 
$3000\,$K for $r_s=1.86$,
around $2500\,$K for $r_s=1.75$, and at approximately $1250\,$K for $r_s=1.6$. 
The maximum of $\Delta P_{mix}/P$ is located at a slightly higher temperature
than the transition from a pure molecular to a mainly atomic phase in pure
hydrogen. 
The linear
mixing rule transfers the behavior of pure hydrogen in an incorrect way into the
mixture. This causes these deviations of up to approximately $15\%$. In 
addition, it is
shown that linear mixing is not a good approximation for hydrogen-helium systems 
containing atoms {\it and} molecules. 
For higher temperatures the deviation from
linear mixing declines although in the considered range it does not reach
values below $5\%$ again.

\begin{figure}[!]
\includegraphics[angle=0,width=\figurewidth]{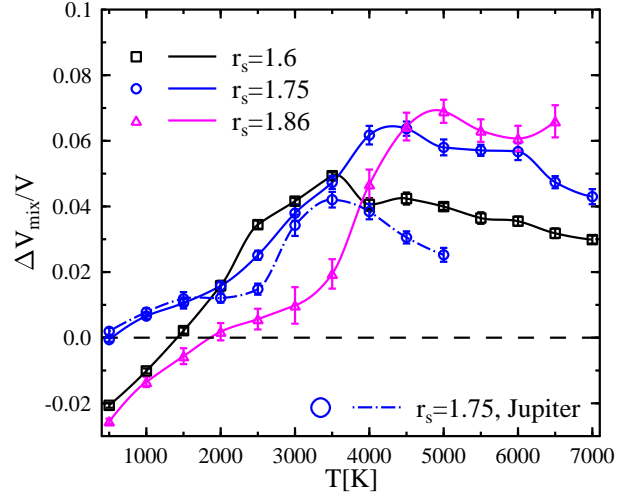}
\caption{Mixing error in the volume due to the linear mixing approximation
at constant pressure for hydrogen-helium for various densities and two 
different mixing ratios of $x=0.5$ (solid lines) and $x=0.14$ (dashed line)
as function of temperature.
}
\label{mixp1}
\end{figure}
A similar statement can be made for mixing at constant pressure as shown in
Fig.\ref{mixp1}. 
The same features as in Fig. \ref{mixv1} can be observed. The maximum
of the mixing error is shifted to lower temperatures for higher densities.
However, the error in the volume introduced by LM is slightly smaller than the 
one in the pressure.
\begin{figure}[!]
\includegraphics[angle=0,width=\figurewidth]{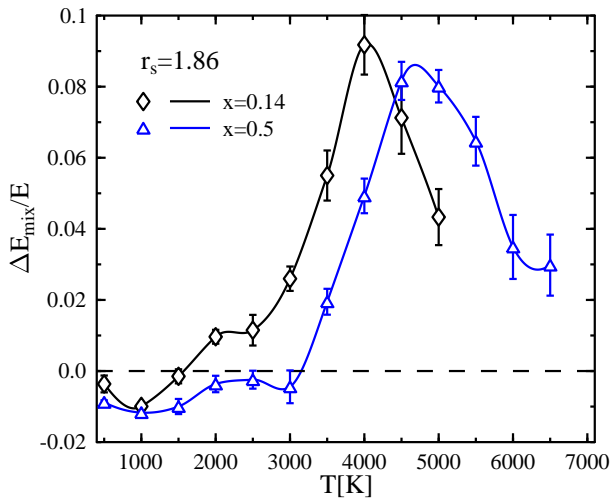}
\caption{Mixing error in the energy due to the linear mixing approximation at 
constant pressure as function of the temperature for hydrogen-helium mixtures
at a density of $r_s=1.86$. The mixing ratios are $x=0.5$ and $x=0.14$ (Jupiter).
 }
\label{mixp2}
\end{figure}
Comparing curves with different mixing ratios at constant density in Figs.
\ref{mixp1} and \ref{mixp2} it is obvious that the maximum in the error is
reached at lower temperatures for smaller mixing ratios $x$. This is in agreement
with the temperature shift of molecular dissociation as function of the helium
ratio in the system. Less obvious is the
actual absolute value for the deviation from linear mixing. Whereas the error in
the volume never exceeds $5\%$, the energy is much more sensitive to deviations
from LM with an error of up to $9\%$.

All, figures \ref{mixv1}, \ref{mixp1}, and \ref{mixp2} show that LM can be 
considered a good
approximation only for systems at low temperatures ($T<1000K$) and at
very high temperatures $T>8000K$ for the densities presented here. In the 
case of liquid hydrogen-helium mixtures, these systems consist of weakly
interacting hydrogen molecules and helium atoms (low $T$) or of weakly interacting
hydrogen and helium atoms (high $T$). More complicated situations where atoms
{\it and} molecules
of hydrogen and helium are involved and interactions between them are 
non negligible require a better description than LM since helium has a 
significant
influence on the dissociation degree and the mixture cannot be considered to be a
composition of two fluids.

The figures presented here could
suggest that LM is a rather good approximation for all of the higher temperatures.
This is by no means the case. As stated before, LM works well only
for nearly ideal systems. When
the temperature becomes too high so that the gas of hydrogen and helium atoms
experiences ionization (this can be accomplished by increasing the density 
as well) and a
partially ionized plasma is created, the merely short ranged interatomic
interactions are replaced by long range Coulomb forces and nonideality
contributions to the EOS  become very important again. In this regime, LM breaks
down again, as was demonstrated by Militzer by means of PIMC \cite{hheburkhard}.
\section{Summary}\label{summ}
We use first principle DFT-MD simulations to study equilibrium properties of
hydrogen and hydrogen-helium mixtures under extreme conditions. The results
obtained are relevant for the modeling of giant gas planets and for the principle
understanding of the EOS of fluid hydrogen and
hydrogen-helium mixtures. 

Our results for pure hydrogen show a smooth transition from a molecular to an 
atomic state which is accompanied by a transition from an insulating to a
metallic like state. In the transition region, we find a negative temperature
derivative of the pressure.
The results for the hydrogen EOS show deviations from
widely used chemical models (up to $20\%$). In particular, the point of 
dissociation for the molecules is obtained at much lower temperatures than in 
chemical models. We find satisfying agreement with previous DFT-MD
simulations only.

In particular, we demonstrate the influence of helium 
on hydrogen molecules. The presence of helium results into more stable molecules 
and an altered transition from a molecular to an atomic fluid state.
Helium reduces the negative slope of the pressure isochores in the transition
region.
The bond length of the hydrogen molecules is shortened by $6\%$ for $x=0.5$. 
As a result, the degree of dissociation is lowered and the electronic bandgap is
increased
The effect of
helium is found to be more important for higher densities where the
stronger localization of the electrons prevents
degeneracy effects for the electrons to become dominant.

Our analysis of the mixing properties for a $x=0.5$ mixture of hydrogen and
helium shows that the corrections to the linear mixing approximation are
significant. Maximum EOS corrections of 15\% were found for mixing at constant
volume and 8\% for mixing at constant pressure. For Jupiter like conditions,
corrections up to 5\% were obtained.

The presented results and forthcoming work should help to clarify 
long standing questions concerning the formation process of giant gas
planets, help restrict the core size of Jupiter, and allow one to make 
predictions for the hydrogen-helium phase separation.
\begin{acknowledgments}
Fruitful discussions with B. Hubbard are acknowledged. 
This material is based upon work supported by NASA under the grant NNG05GH29G,
by the Carnegie Institution of Canada, and by the NSF under the
grant 0507321. I.T. and S.A.B. acknowledge support by the NSERC of Canada.
\end{acknowledgments}
%
%
\bibliographystyle{unsrt}

\begin{thebibliography}{10}

\bibitem{mayor95}
M.~Mayor and D.~Queloz.
\newblock {\em Nature}, 378:355, 1995.

\bibitem{Bu05}
A.~Burrows.
\newblock {\em Nature}, 433:261, 2005.

\bibitem{exopl}
The extrasolar planet encyclopedia.
\newblock http://exoplanet.eu.

\bibitem{Gu02}
T. Guillot {\it el al.}, in {\it Jupiter}, F. Bagenal, Ed. (Univ. of Arizona
  Press, Tucson, 2003), chap. 3, pp. 35-57.

\bibitem{SG04}
D.~Saumon and T.~Guillot.
\newblock {\em Astrophys. J.}, 609:1170, 2004.

\bibitem{Stevenson75}
D.J. Stevenson.
\newblock {\em Phys. Rev. B}, 12:3999, 1975.

\bibitem{SC92}
D.~Saumon and G.~Chabrier.
\newblock {\em Phys. Rev. A}, {46}:2084, 1992.

\bibitem{SC95}
D.~Saumon, G.~Chabrier, and H.~M.~Van Horn.
\newblock {\em Astrophys. J. Suppl.}, 99:713, 1995.

\bibitem{WC05}
C.~Winisdoerffer and G.~Chabrier.
\newblock {\em Phys. Rev. E}, 71:026402, 2005.

\bibitem{Ju01}
H.~Juranek, R.~Redmer, and Y.~Rosenfeld.
\newblock {\em J. Chem. Phys.}, 117:1768, 2002.

\bibitem{ju03}
H.~Juranek, V.~Schwarz, and R.~Redmer.
\newblock {\em J. Phys A}, 36:6181, 2003.

\bibitem{bez04a}
V.~Bezkrovniy, M.~Schlanges, D.~Kremp, and W.-D. Kraeft.
\newblock {\em Phys. Rev. E}, 69:061204, 2004.

\bibitem{Be99}
D.~Beule, W.~Ebeling, A.~F{\"o}rster, H.~Juranek, S.~Nagel, R.~Redmer, and
  G.~R\"opke.
\newblock {\em Phys. Rev. B}, 59:14177, 1999.

\bibitem{Ro98}
M.~Ross.
\newblock {\em Phys. Rev. B}, {58}:669, 1998.

\bibitem{KS65}
W.~Kohn and L.J. Sham.
\newblock {\em Phys. Rev.}, {140}:A1133, 1965.

\bibitem{Klepeis1991}
J.~E. Klepeis, K.~J. Schafer, T.~W.~Barbee III, and M.~Ross.
\newblock {\em Science}, 254:986, 1991.

\bibitem{mazin95}
I.I. Mazin and R.~E. Cohen.
\newblock {\em Phys. Rev. B}, 52:R8597, 1995.

\bibitem{CP85}
R.~Car and M.~Parrinello.
\newblock {\em Phys., Rev. Lett.}, {55}:2471, 1985.

\bibitem{winis04}
C.~Winisdoerffer, G.~Chabrier, and G.~Z\`erah.
\newblock {\em Phys. Rev. E}, 70:026403, 2004.

\bibitem{ko95}
J.~Kohanoff and J.-P. Hansen.
\newblock {\em Phys. Rev. E}, 54:768, 1996.

\bibitem{DW02}
M.~W.~C. Dharma-wardana and F.~Perrot.
\newblock {\em Phys. Rev. B}, 66:014110, 2002.

\bibitem{dw82}
N.W.C. Dharma-wardana and F.~Perrot.
\newblock {\em Phys. Rev. A}, 26:2096, 1982.

\bibitem{xu98}
H.~Xu and J.-P. Hansen.
\newblock {\em Phys. Rev. E}, 57:211, 1998.

\bibitem{Ma96}
W.~R. Magro, D.~M. Ceperley, C.~Pierleoni, and B.~Bernu.
\newblock {\em Phys. Rev. Lett.}, {76}:1240, 1996.

\bibitem{PC94}
C.~Pierleoni, D.M. Ceperley, B.~Bernu, and W.R. Magro.
\newblock {\em Phys. Rev. Lett.}, {73}:2145, 1994.

\bibitem{Pierleoni04}
C.~Pierleoni, David~M. Ceperley, and M.~Holzmann.
\newblock {\em Phys. Rev. Lett.}, 93:146402, 2004.

\bibitem{tsch}
M.~Schlanges, M.~Bonitz, and A.~Tschttschjan.
\newblock {\em Contr. Plasma Phys.}, 35:109, 1995.

\bibitem{vorb04}
J.~Vorberger, M.~Schlanges, and W.-D. Kraeft.
\newblock {\em Phys. Rev. E}, 69:046407, 2004.

\bibitem{MC00}
B.~Militzer and D.~M. Ceperley.
\newblock {\em Phys. Rev. Lett.}, 85:1890, 2000.

\bibitem{Mi01}
B.~Militzer {\it et al.}
\newblock {\em Phys. Rev. Lett.}, 87:275502, 2001.

\bibitem{bez04}
V.~Bezkrovniy, V.S. Filinov, D.~Kremp, M.~Bonitz, M.~Schlanges, W.-D. Kraeft,
  P.R. Levashov, and V.E. Fortov.
\newblock {\em Phys. Rev. E}, 70:057401, 2004.

\bibitem{bonev2004}
S.~A. Bonev, B.~Militzer, and G.~Galli.
\newblock {\em Phys. Rev. B}, 69:014101, 2004.

\bibitem{Mi06}
B.~Militzer.
\newblock {\em submitted to Phys. Rev. Lett.}, 2006.

\bibitem{Scandolo2003}
S.~Scandolo.
\newblock {\em Proc. Nat. Ac. Sci.}, 100:3051, 2003.

\bibitem{pfaffenzeller97}
O.~Pfaffenzeller and D.~Hohl.
\newblock {\em J. Phys.: Condens. Matter}, 9:11023, 1997.

\bibitem{bonevNature}
S.A. Bonev, E.~Schwegler, T.~Ogitsu, and G.~Galli.
\newblock {\em Nature}, 431:669, 2004.

\bibitem{natoli95}
V.~Natoli, R.M. Martin, and D.~Ceperley.
\newblock {\em Phys, Rev, Lett.}, 74:1601, 1995.

\bibitem{johnson00}
K.A. Johnson and N.W. Ashcroft.
\newblock {\em Nature}, 403:632, 2000.

\bibitem{ki00}
H.~Kitamura, S.~Tsuneyuki, T.~Ogitsu, and T.~Miyake.
\newblock {\em Nature}, {}, 2000.

\bibitem{Wi35}
E.~Wigner and H.~B. Huntington.
\newblock {\em J. Chem. Phys.}, {3}:764, 1935.

\bibitem{Jo96}
M.D. Jones and D.M. Ceperley.
\newblock {\em Phys. Rev. Lett.}, 76:4572, 1996.

\bibitem{MG05}
B.~Militzer and R.~L. Graham.
\newblock {\em Journal of Physics and Chemistry of Solids}, submitted (2005).

\bibitem{Pfaffenzeller1995}
O.~Pfaffenzeller, D.~Hohl, and P.~Ballone.
\newblock {\em Phys. Rev. Lett.}, 74:2599, 1995.

\bibitem{hheburkhard}
B.~Militzer.
\newblock {\em J. Low. Temp. Phys}, 139:739, 2005.

\bibitem{CPMD}
CPMD, Copyright IBM Corp 1990-2006, MPI f{\"ur} Festk{\"o}rperforschung
  Stuttgart 1997-2001.

\bibitem{PBE}
J.~P. Perdew, K.~Burke, and M.~Ernzerhof.
\newblock {\em Phys. Rev. Lett.}, 77:3865, 1996.

\bibitem{TM91}
N.~Troullier and J.~L. Martin.
\newblock {\em Phys. Rev. B}, 43:1993, 1001.

\bibitem{fhi}
M.~Fuchs and M.~Scheffler.
\newblock {\em Comp. Phys. Com.}, 119:67, 1999.

\bibitem{Abinit}
Copyright ABINIT group (M. Mikami, J. M. Beuken, X. Gonze) 2003-2005.

\bibitem{MP76}
H.J. Monkhorst and J.D. Pack.
\newblock {\em Phys. Rev. B.}, 13:5188, 1976.

\bibitem{Lo02}
P.~Loubeyre, F.~Occelli, and R.~LeToullec.
\newblock {\em Nature}, 416:613, 2002.

\bibitem{CA87}
D.M. Ceperley and B.J. Alder.
\newblock {\em Phys. Rev. B}, {36}:2092, 1987.

\bibitem{We96}
S.T. Weir, A.C. Mitchell, and W.J. Nellis.
\newblock {\em Phys. Rev. Lett.}, 76:1860, 1996.

\bibitem{gi84}
P.~Giannozzi and S.~Baroni.
\newblock {\em Phys. Rev. B}, 30:7187, 1984.

\bibitem{HoodGalli}
R.~J. Hood and G.~Galli.
\newblock {\em J. Chem. Phys.}, 120:5691, 2004.

\bibitem{Ju00}
H.~Juranek and R.~Redmer.
\newblock {\em J. Chem. Phys.}, 112:3780, 2000.

\bibitem{KRT98}
M.~Knaup, P.-G. Reinhard, and C.~Toepffer.
\newblock {\em Contrib. Plasma Phys.}, {39 }1-2:57, 1999.

\bibitem{Co95}
L.~Collins, I~Kwon, J.~Kress, N.~Troullier, and D.~Lynch.
\newblock {\em Phys. Rev. E}, {52}:6202, 1995.

\bibitem{Lou96}
P.~Loubeyre, R.~LeToullec, D.~Hausermann, M.~Hanfland, R.J. Hemley, H.K. Mao,
  and L.W. Finger.
\newblock {\em Nature}, 383:702, 1996.

\bibitem{Kn02}
M.~Knaup.
\newblock PhD thesis, University of Erlangen, Germany, 2002.

\bibitem{Kna02}
M.~Knaup, P.-G. Reinhardt, C.~Toepffer, and G.~Zwicknagel.
\newblock {\em Comp. Phys. Com.}, 147:202, 2002.

\bibitem{Stevenson77a}
D.J. Stevenson and E.E. Salpeter.
\newblock {\em Astrophys. J. Suppl. Ser.}, 35:221, 1977.

\bibitem{Stevenson77b}
D.J. Stevenson and E.E. Salpeter.
\newblock {\em Astrophys. J. Suppl.}, 35:239, 1977.

\bibitem{SCapj92}
G.~Chabrier, D.~Saumon, W.B. Hubbard, and J.I. Lunine.
\newblock {\em Astrophys. J.}, 391:817, 1992.

\bibitem{Hu85}
W.B. Hubbard and H.E. deWitt.
\newblock {\em Astrophys. J.}, 290:388, 1985.

\bibitem{Og93}
S.~Ogata, H.~Iyetomi, S.~Ichimaru, and H.M.~Van Horn.
\newblock {\em Phys. Rev. E}, 48:1344, 1993.

\bibitem{DeW96}
H.E. DeWitt, W.~Slattery, and G.~Chabrier.
\newblock {\em physica B}, 228:21, 1996.

\bibitem{Rosen96}
Y.~Rosenfeld.
\newblock {\em Phys. Rev. E}, 54:2827, 1996.

\bibitem{DeW03}
H.E. DeWitt and W.~Slattery.
\newblock {\em Contrib. Plasma Phys.}, 43:279, 2003.

\bibitem{DeW93b}
H.E. DeWitt.
\newblock {\em Equation of State in Astrophysics}, page 330.
\newblock IAU Colloquium 147. Cambridge University Press, 1994.

\end{thebibliography}

%
\end{document}